\documentstyle[12pt,epsf]{article}

\newcommand{\be}{\begin{equation}}
\newcommand{\ee}{\end{equation}}

\begin{document}

\title {Dynamic Behavior of Spin Glass Systems on Quenched 
$\phi^{3}$ Graphs}

\author{C. Baillie$^{(a)}$, D. A. Johnston$^{(b)}$, E. Marinari$^{(c)}$\\
   and C. Naitza$^{(c)}$\\[0.5em]
  {\small (a): Dept. of Computer Science, University of Colorado,}\\
  {\small \ \  Boulder, CO 80309  (USA)}\\
  {\small (b): Department of Mathematics, Heriot-Watt University,}\\
  {\small \ \ Edinburgh, EH14 4AS, Scotland (UK)}\\
  {\small (c): Dipartimento di Fisica and Infn, Universit\`a di Cagliari}\\
  {\small \ \  Via Ospedale 72, 07100 Cagliari (Italy)}\\[0.3em]
  {\small \tt clive@kilt.cs.colorado.edu}\\
  {\small \tt des@ma.hw.ac.uk}\\
  {\small \tt marinari,naitza@ca.infn.it}\\[0.5em]}


\maketitle
\begin{abstract}
We study the dynamical out-of-equilibrium behavior of a $J = \pm 1$
Ising spin glass on quenched $\phi^{3}$ graphs. We show that magnetization 
and energy decay with a power law behavior, with exponents that are 
linear in $\frac{T}{T_{c}}$. Quenched $\phi^{3}$ graphs turn out to 
be a very effective way to study numerically mean field spin glasses.
\end{abstract}
\vfill
\begin{flushright}
  { \tt cond-mat/9606194 }
\end{flushright}

\newpage

Spin models on quenched $\phi^{3}$ graphs have been considered in the last 
few years as a possible effective way to define mean field models
\cite{ONE,TWO}. In general one can consider $\phi^{z}$ random quenched 
graphs (with the limit $z\to \infty$ coinciding with the usual 
Sherrington-Kirkpatrick definition of the mean field).  One obtains 
mean field behavior on such graphs for the same reason as on the Bethe 
lattice because of the tree-like local structure, but the drawback of 
a dominant surface contribution is absent.  It is not clear whether 
this approach is advantageous analytically since the large number law 
cannot be applied in a straightforward manner, and in general 
calculations appear more involved than for the SK 
approach\footnote{The saddle point equations for replica Ising spins 
can, however, be cast in a rather appealing form \cite{TWO}.}.  On the 
other hand from the numeric point of view there {\it are} distinct 
advantages, since in this case one is dealing with a model with local 
interaction where one spin update takes the order of $z$ operations 
and not order volume.  A parenthetic warning is probably in order at 
this point - the fact that computation is faster does not necessarily 
imply that the model is the best choice.  One has to perform the 
simulation to check.  For example, the hypercube definition of the 
mean field (see for example Parisi-Ritort in \cite{ONE}) undergoes 
very strong finite size effects, that can make the analysis of the 
thermodynamic limit very obscure.

Gardner, Derrida and Mottishaw \cite{GADEMO} have drawn
attention to the fact that when looking at the dynamical behavior of 
disordered systems one can expect to see power law behavior. The complex 
landscape does not allow a fast exponential convergence. Their 
calculation (that integrates exactly a small number of time steps) 
establishes that the $T=0$ parallel dynamics from a magnetized state 
leads to a non-zero magnetization expectation value with a power 
correction. Eisfeller and Opper in ref. \cite{EISOPP} introduced a new 
approach combining the dynamical functional method and a Monte 
Carlo simulation of a stochastic single-spin equation that allowed 
the determination with good precision of the value of the remanent 
magnetization, $m(\infty)=0.474$ and of the exponent 
$\mu(T=0)=0.184$ 
where
\begin{equation}
  \protect\label{E-ONE}
  m(t)-m(\infty)\simeq (t/\tau)^{-\mu}.
\end{equation}

In a recent work Ferraro\cite{FERRAR} has generalized the 
Eisfeller-Opper work to the non-zero temperature case.  In this way he 
has been able to establish that in the SK model the magnetization and 
the energy do decay with power laws.  Such a generalization of the 
Eisfeller-Opper method to non-zero $T$ does indeed allow a good 
determination of the magnetization exponent (which turns out to be 
linear in $T/T_{c}$), but is less effective for the energy exponent.  
Very large scale dynamical simulations of the SK model from Rossetti 
\cite{ROSSET} allowed a first direct Monte Carlo determination of the 
power exponents, and showed that in the case of the SK definition of 
the mean field numerical simulation is very tough due to the non-local 
interaction.  For comparison it is worth noting what happens in real 
experiments with typical spin glasses (see for example \cite{SOU}).  
If one fits with the form in equation (\ref{E-ONE}) one finds that 
$\mu(T)\simeq 0.4 T/T_{c}$, where $\tau$ is of the order of magnitude 
of the microscopic time for single spin flip, and turns out not to 
depend on $T$.

Given the difficulty of simulations with non-local interactions and 
the success of the $\phi^3$ graph simulations in reproducing static 
mean-field spin glass behavior \cite{TWO} it is obviously tempting to 
investigate dynamical aspects of spin glass behavior with $\phi^3$ 
graphs.  We have thus carried out a Monte Carlo simulation of the 
dynamics of a spin glass model on $\phi^{3}$ graphs for different 
temperature values and graph sizes.  The principal measurements of 
interest in such a simulation are the time series of the energy and 
the magnetization, which we have measured in the standard fashion.  
From these time series we have systematically analyzed the time 
dependence of magnetization and energy, starting from a cold system 
with $m=1$.  We have tried to find the temporal regions where we could 
exhibit a clean scaling behavior, and we discuss our findings in the 
following.

We have used integer uniformly distributed quenched random couplings
$J=\pm 1$ (i.e. the probability
distribution for the quenched bond distribution is
$P(J) = 1/2 \delta (J+1) + 1/2\delta(J-1)$)
in our simulations here
because the earlier static simulations carried
out with this distribution were known to
give convincing agreement with theoretical calculations
\cite{TWO}.
This provided a degree of confidence in the code used.
One slight drawback of this choice
is that with integer couplings the system has a energy gap, and at too 
low $T$ values the system will not be able to converge to equilibrium 
(as we have checked in preliminary numerical simulations). In 
order to avoid this problem we have kept our $T$ values 
sufficiently high to avoid the 
influence of the gap.
In any case, preliminary work with a gaussian coupling distribution
where there is no gap
indicates no fundamental differences with the results here.

For each of the graph sizes simulated 100 different 
$\phi^3$ graphs were generated and a 
massively parallel processor (the Intel Paragon)
was employed to allow the quenched averaging
to be performed {\it in situ}. As we are 
interested in the ``real'' dynamics of the model
we employed a simple single spin Metropolis
update. The actual runs were of relatively
short duration ($O(10^4)$ sweeps) as this was
sufficient to cover the temporal regions of interest.
As we have indicated above, a cold start with $m=1$
was used. 

Having dealt with the preliminaries, we
now discuss the magnetization data
in general terms. We start 
from a magnetized system, and observe the magnetization decay. On 
general grounds, we expect three different temporal regimes. At short 
times there is a transient, non-universal region, which is expected 
to depend on the details of the dynamics (in our case Metropolis).
For intermediate times we expect to be able to 
detect a region with time power decay, where 

\begin{equation}
  m(t) \simeq \frac{A}{t^{\mu}}\ .
\end{equation}
Finally, at large times on a finite graph we reach a plateau value for the 
remanent magnetization. The decay rate to this plateau (when, due to the 
finite extent of the sample, the system has reached the bottom of a 
valley) is a new dynamical effect, that can again be different from 
the previous phase. We will see that it is probably an exponential 
decay to the bottom of the hole. As we already have observed, in a 
finite size system of volume $V$ one expects 

\begin{equation}
m(t) \to_{t\to\infty} m_{(V)}\ .
\end{equation}
In fact our data is compatible with
a zero infinite volume limit for $m_{(V)}$

\begin{equation}
m_{(V)} \to_{V\to\infty} 0\ ,
\end{equation}
though other possibilities are not excluded within the
accuracy of the measurements.

\begin{figure}[bhp]
  \epsfxsize=400pt\epsffile{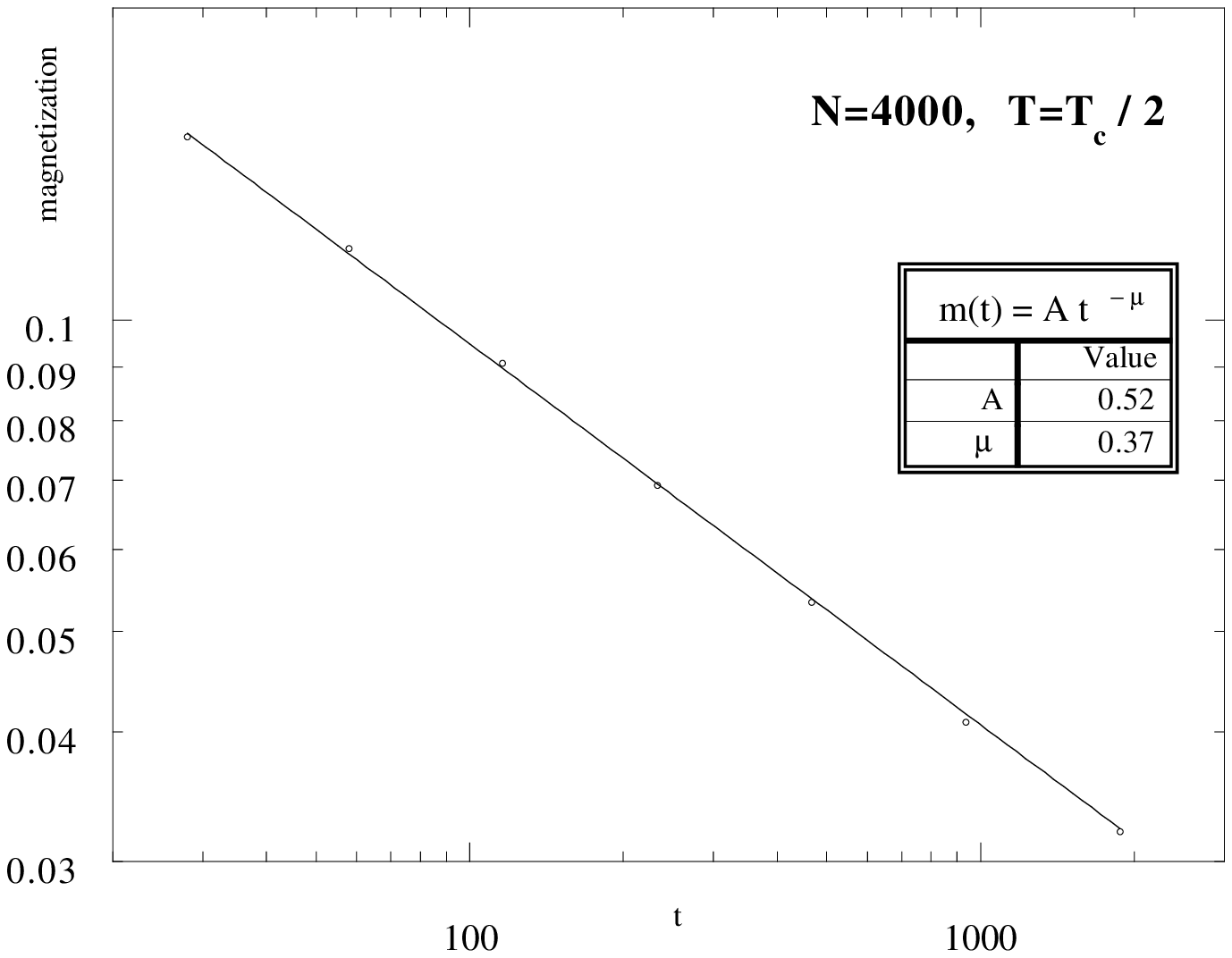}
  \caption[1]{
    Magnetization versus Monte Carlo time in log-log scale. Here
    $N=4000$, $T=\frac{T_{c}}{2}$, $t$ ranging from $20$ to $2500$.
  }
  \protect\label{F-FIG1}
\end{figure}

Let us now discuss in some detail the analysis of the 
magnetization data on the largest graphs ($N=4000$ vertices)
simulated. For this graph 
size we have data samples of $20,000$ lattice sweeps at each 
$\beta$ value.
The coldest data we will discuss is at $\beta=2\beta_{c}$ 
($\beta_{c} \simeq .881$ for
the $J=\pm 1$ distribution on $\phi^3$ graphs). For lower 
temperature values the discreteness of our couplings starts to play a 
role, and it is difficult to be sure of having determined the 
asymptotic power behavior. In fig. (\ref{F-FIG1}) we plot the data 
we use for our best fit. We use here times from $t_{min}=20$ 
to $t_{max}=2500$. Our best fit, drawn in the figure, gives an 
exponent $\mu=0.37$. This fit is very stable. Repeating it by 
doubling the confidence window both at small times and at large times 
(i.e. by selecting $t_{min}=40$  and $t_{max}=1250$) we find 
$\mu=0.37$.

The magnetization data at $\beta=2\beta_{c}$ converges to a plateau in 
our large time region (i.e. $t\simeq 15000$). A fit to a single power 
converging to a constant plateau does not work. In order to make the 
fit work one has to add higher power terms or an exponential decay 
(see later). The physics of what is happening is quite clear: we have 
a power decay for intermediate times, and at large times the finite 
volume system reaches the finite size value of the 
magnetization. The decay to such a finite value is not governed by the 
asymptotic power law we measure in the intermediate time region, and 
is probably (but not certainly) exponential.

\begin{figure}[bhp]
  \epsfxsize=400pt\epsffile{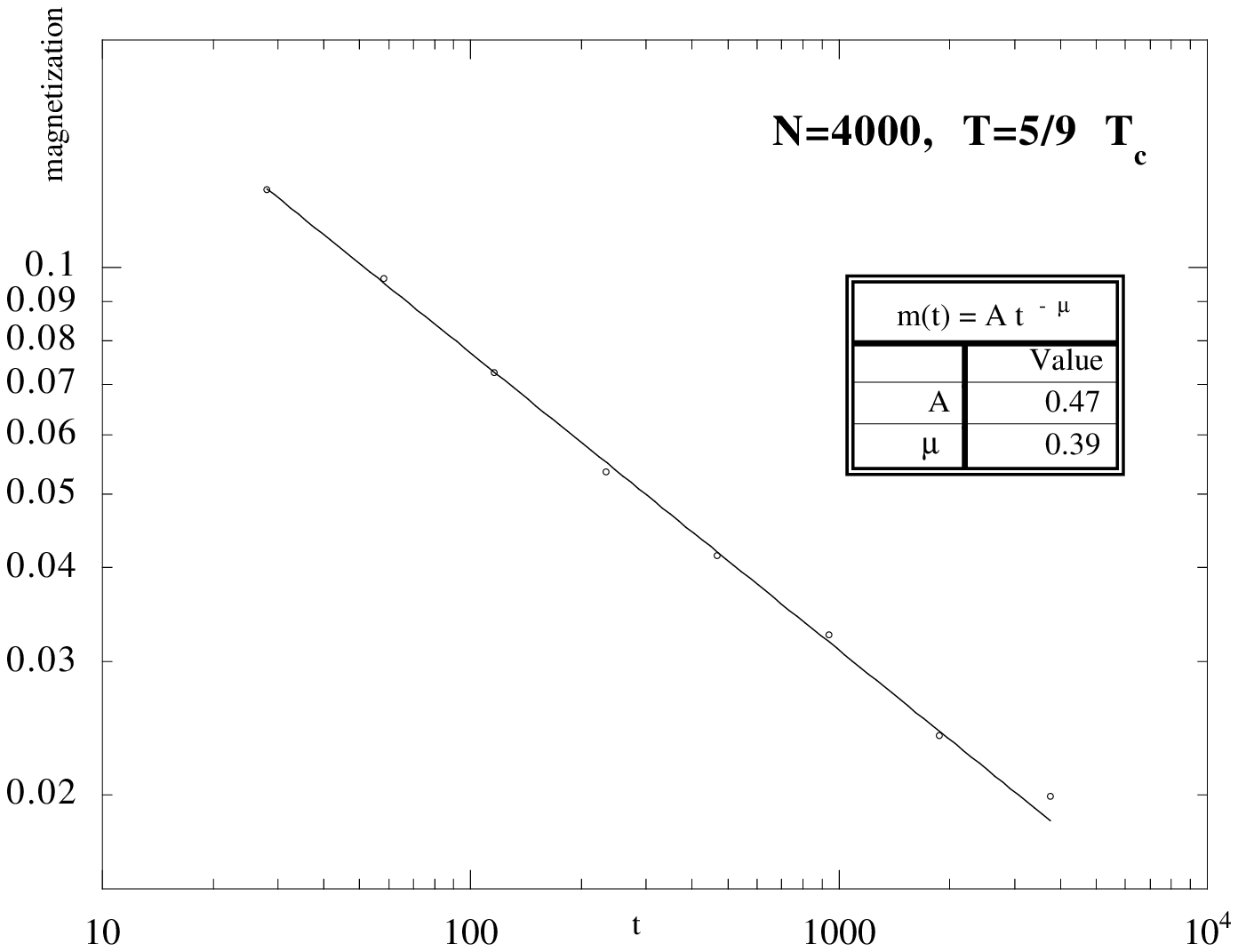}
  \caption[1]{
    Magnetization versus Monte Carlo time in log-log scale. Here
    $N=4000$, $T=\frac{5T_{c}}{9}$, $t$ ranging from $20$ to $2500$.
  }
  \protect\label{F-FIG2}
\end{figure}

\begin{figure}[bhp]
  \epsfxsize=400pt\epsffile{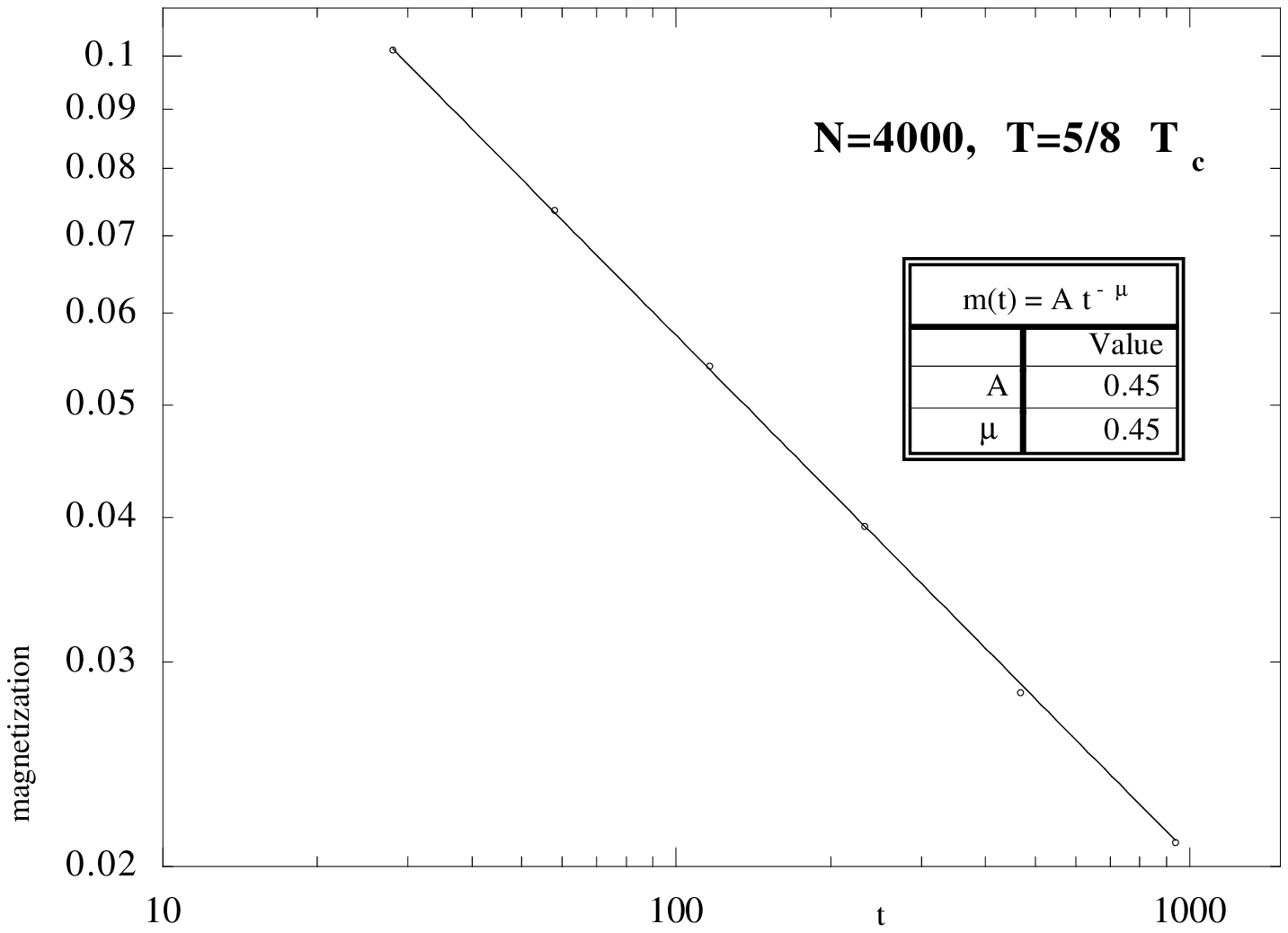}
  \caption[1]{
    Magnetization versus Monte Carlo time in log-log scale. Here
    $N=4000$, $T=\frac{5T_{c}}{8}$, $t$ ranging from $40$ to $1250$.
  }
  \protect\label{F-FIG3}
\end{figure}

In fig. (\ref{F-FIG2}) we plot the magnetization data at 
$\beta=\frac95\beta_{c}$.
Here the last point is already slightly off a good power fit 
because the 
time needed to reach the plateau is larger for lower $T$ values. We 
report this fit in order to give a feeling of the kind of systematic 
effects one gets, and because the fit with $t_{max}=1250$ gives for 
the first two significant digits of the exponent the same result. 
Again, the fit is stable.
The fit in fig. (\ref{F-FIG3}), at $\beta=\frac85\beta_{c}$, uses 
only time points ranging from $40$ to $1250$, but its quality, with an 
exponent of $\mu=.45$, is again very good.

\begin{figure}[bhp]
  \epsfxsize=400pt\epsffile{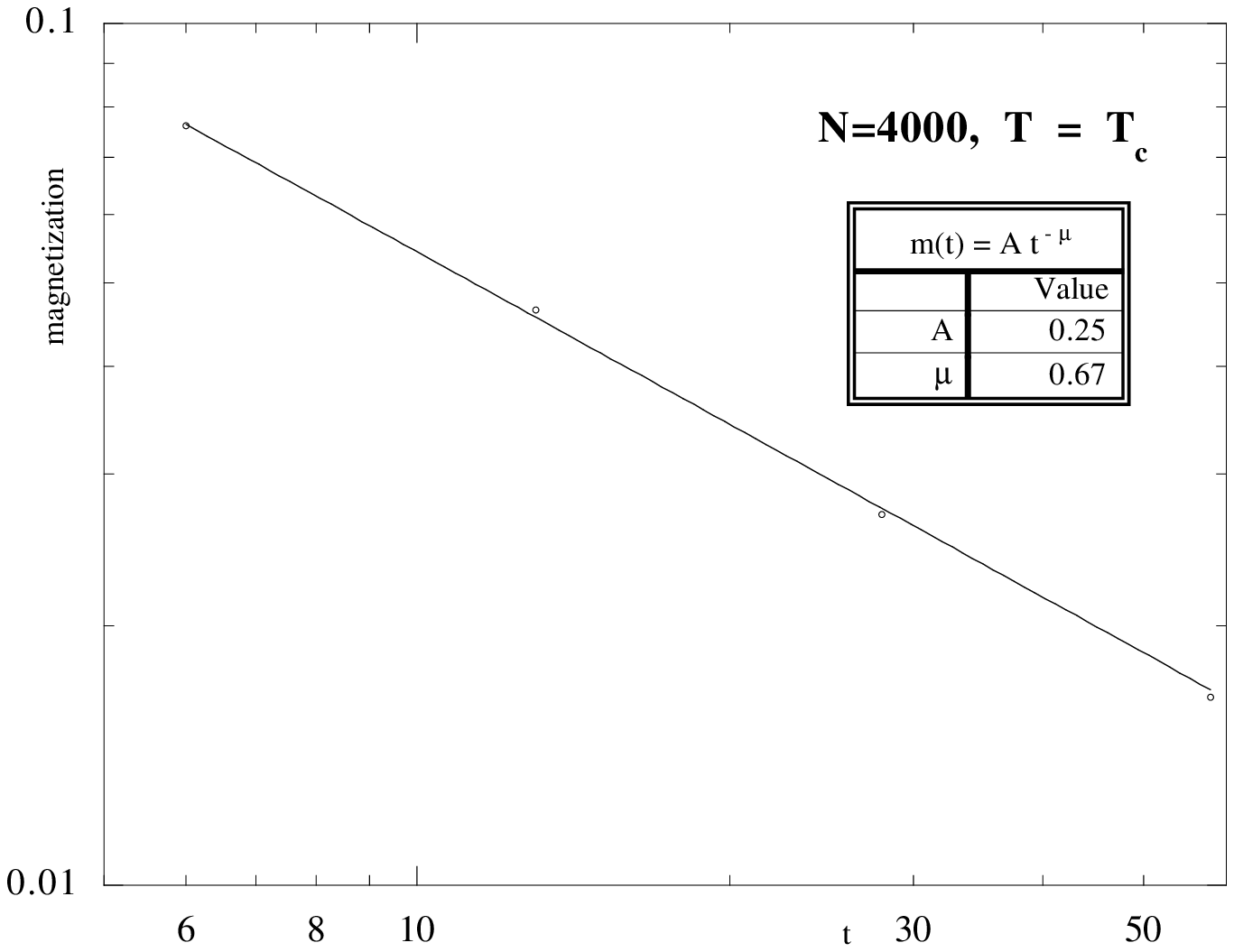}
  \caption[1]{
    Magnetization versus Monte Carlo time in log-log scale. Here
    $N=4000$, $T=T_{c}$, $t$ ranging from $4$ to $77$.
  }
  \protect\label{F-FIG4}
\end{figure}

\begin{figure}[bhp]
  \epsfxsize=400pt\epsffile{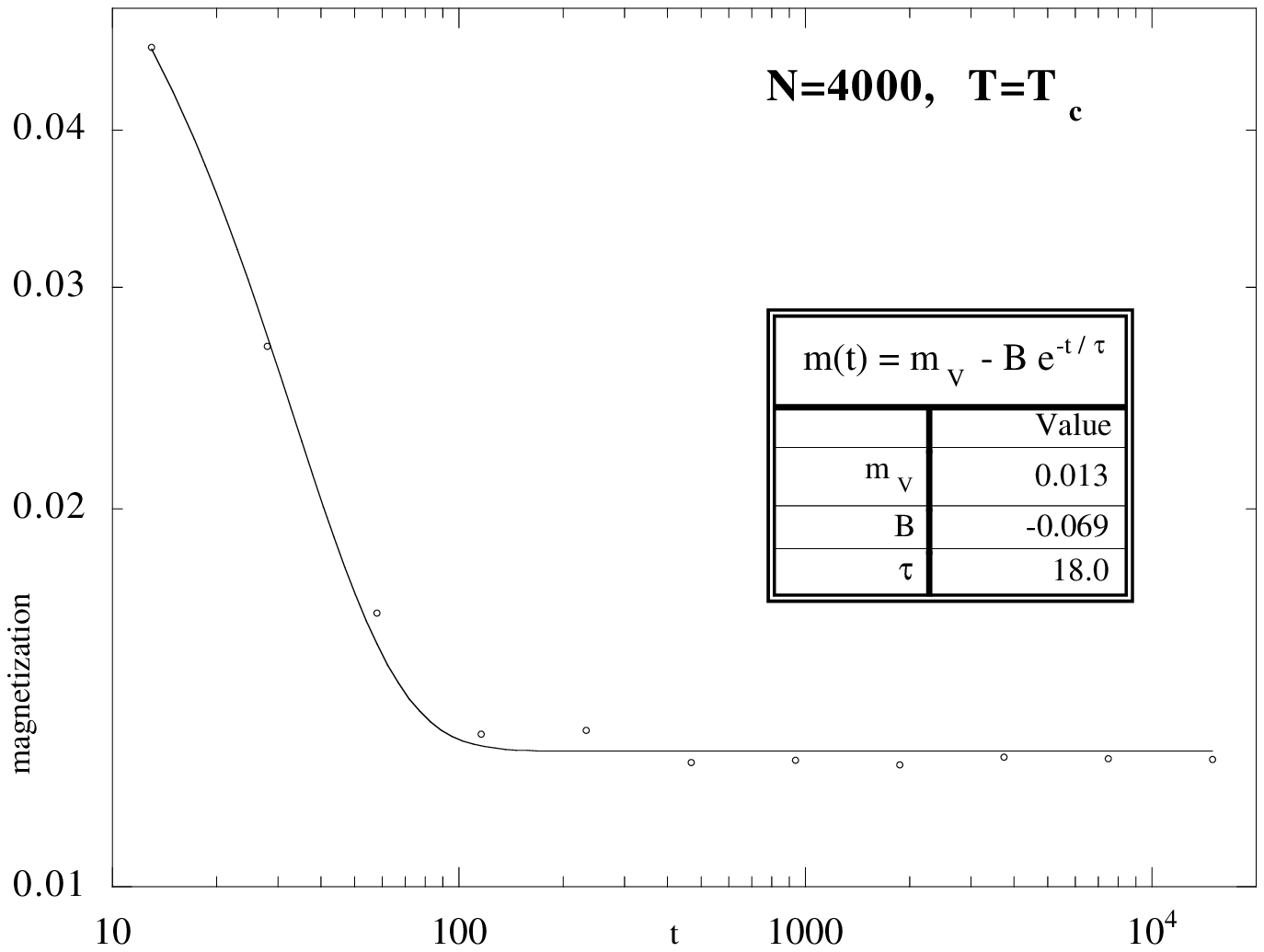}
  \caption[1]{
    Magnetization versus Monte Carlo time in log-log scale. Here
    $N=4000$, $T=T_{c}$, $t$ ranging from $9$ to $20000$.
  }
  \protect\label{F-FIG5}
\end{figure}

Power fits for values of $T$ closer to $T_{c}$ are equally good, and 
we report their results in Table (\ref{T-TAB1}). The time window we use 
moves to short times when $T\to T_{c}$ as can be seen
in fig. (\ref{F-FIG4}) for the time series at $T=T_{c}$.
At $T_{c}$ the presence of a finite expectation value for $m$ due to 
the finite graph size is very clear. In fig. (\ref{F-FIG5}) we try 
an exponential fit to the decay to the finite size {\em plateau} which 
in this case is reached after only $100$ Monte Carlo sweeps. The 
exponential fit is very good. 
In summary, a power fits well in the  
intermediate time region, while an exponential fit explains very well 
the large time region. A fit with additional,
different power terms is also able to 
fit the large time data: 
we consider our evidence for the existence of such an exponential 
behavior as only qualitative.

\begin{figure}[bhp]
  \epsfxsize=400pt\epsffile{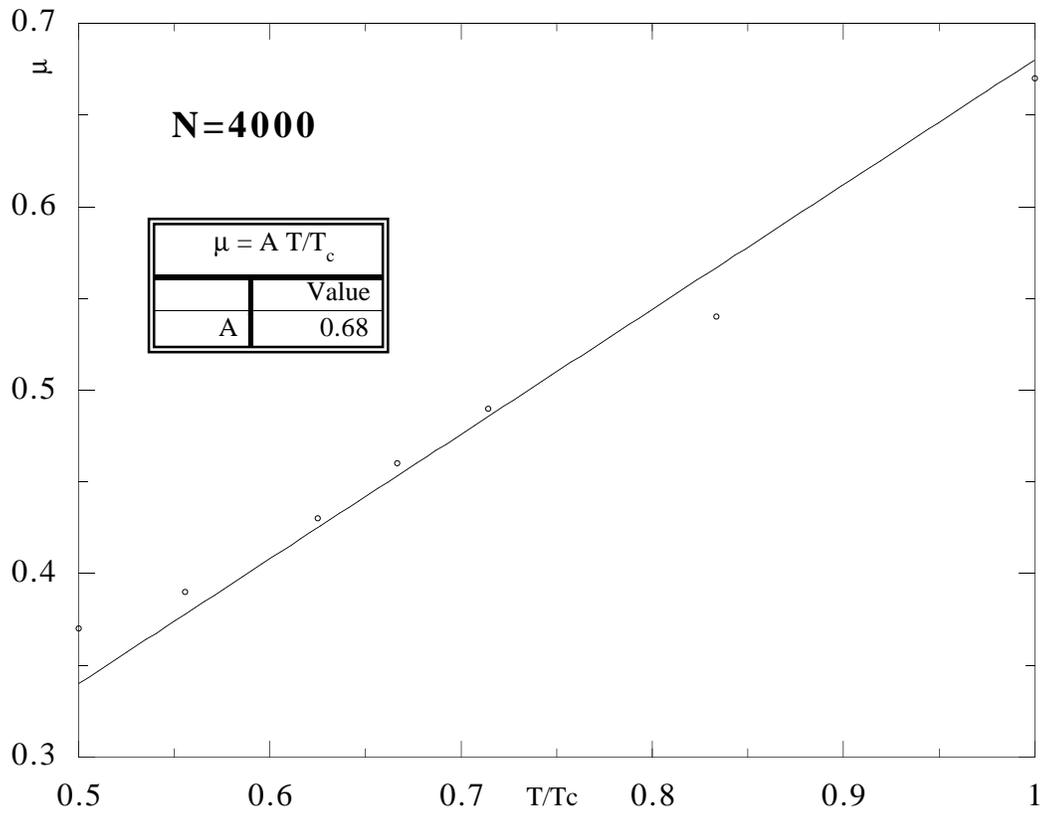}
  \caption[1]{
    The magnetization time decay exponent $\mu$ versus
    $\frac{T}{T_{c}}$ for $N=4000$. The straight line passing through
    the origin is the best fit
    to the data.
  }
  \protect\label{F-FIG6}
\end{figure}

Having established the power law decay at intermediate times,
we now discuss the behavior of the exponent as the 
temperature is varied.
In fig. (\ref{F-FIG6}) we show $\mu(\frac{T}{T_{c}})$.  
The straight line
passing through the origin is our best fit  to the data. The linear 
fit works well from $T_{c}$ down to $0.6$ 
$T_{c}$, while around $\frac12$ $T_{c}$ the fit slightly 
undershoots  the data point. Determining the exponent at low $T$ 
values becomes quite difficult (because it becomes small!) so such a 
small  discrepancy is probably not a problem. Our conclusion is that 

\be
  \mu(T) \simeq \frac23 \frac{T}{T_{c}}\ ,
\ee
in a large $T$ region in the broken phase. The best estimate from both 
Ferraro \cite{FERRAR} and Rossetti \cite{ROSSET}for the magnetization 
exponent is a linear dependence over $T/T_{c}$ with coefficient $1$, 
while, as we have already remarked, experiments favor a value close 
to $.4$. 
By a happy coincidence we find $2/3$, which lies between these two values. 
With the precision allowed by statistical and systematic errors (both in 
numerical simulations and, as far as we can understand, in real 
experiments) this is a very satisfactory result.
Fits of magnetization time series on smaller graphs confirm the 
findings on the $N=4000$ graphs. For example with $N=3000$ we 
estimate $\mu(T_{c}/2)=.36$, and  $\mu(5T_{c}/6)=.55$, in very good 
agreement with the results of table (\ref{T-TAB1}).

We close this section with
two final remarks about the magnetization time series.  Firstly, 
we emphasize again that the 
evidence for a final exponential approach to the plateau value is not 
overwhelmingly compelling, 
being mainly based on data close to $T_{c}$.  Far from 
$T_{c}$ after $20,000$ sweeps we are still quite far from the plateau.  
In all cases fits done by allowing for two or three different powers 
also work quite well.  Secondly the plateau value for the 
magnetization looks quite constant with $T$ in the broken phase.  At 
$T_{c}$ for $N=4000$ we have $m_{\infty}\simeq 0.013$, at $N=3000$ we 
have $m_{\infty}\simeq 0.0145$. The remanent magnetization is becoming 
smaller with increasing volume, but we cannot be sure about the 
absence of a remanent magnetization at non-zero $T$. We stress 
that our measurements of power law do not rely on that, since we have 
selected a scaling region far away from the asymptotic (zero or 
non-zero) limit.

The discussion of the energy exponent  $\epsilon$,

\be
  E(t) \simeq E_{\infty} - \frac{E_{1}}{t^{\epsilon}}
\ee
goes along very similar lines. Here we can fit up to larger times: 
it is only very close to $T_{c}$ that
the power behavior is spoiled, even for
times up to $t=20000$. The difference with the 
behavior of the magnetization, where all of the non-zero plateau at 
$t=\infty$ is probably due to finite size effects, is considerable. At 
$T=\frac{T_{c}}{2}$ fitting from $t=100$ to $t=20000$ we find a stable 
result, and a perfect fit with a very low  $\chi^2$ value. We report 
the best fits to the exponents $\epsilon(T)$ in table (\ref{T-TAB2}).

For lower $\beta$ values we only have to discard a few more Monte 
Carlo points to get a good fit. Only right at $T=T_{c}$ 
does it seem that the asymptotic (non-simple power) behavior is 
encountered too early to get a good 
determination of the critical exponent. 
In this case larger graphs are probably 
needed.

\begin{figure}[bhp]
  \epsfxsize=400pt\epsffile{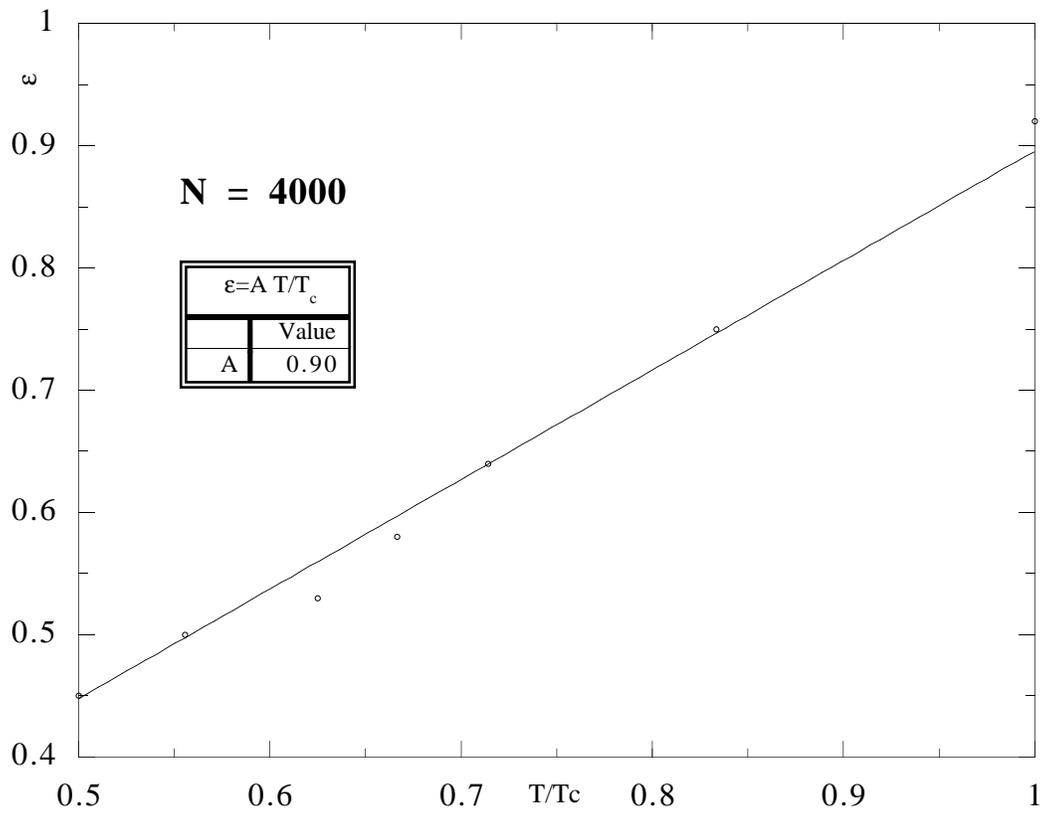}
  \caption[1]{
    The energy time decay exponent $\epsilon$ versus
    $\frac{T}{T_{c}}$ for $N=4000$. The straight line passing through
    the origin is the best fit
    to the data.
  }
  \protect\label{F-FIG7}
\end{figure}

In fig. (\ref{F-FIG7}) we plot our results for $\epsilon$ as a 
function of  $\frac{T}{T_{c}}$, and our best linear fit (that is very 
good). The best fit gives

\be
  \epsilon \simeq  0.9\  \frac{T}{T_{c}}\ .
\ee
We note that this is the first time that it has been possible to get an
estimate of the dependence of the energy exponent on $T$ for 
mean-field like systems.  Results from the large scale simulation 
by Rossetti \cite{ROSSET} for SK model are good but not as clear cut 
as the ones we are able to find here, while the infinite volume 
approach by Eisfeller and Opper \cite{EISOPP} and Ferraro 
\cite{FERRAR} gives clear results for the magnetization exponent but 
not for the energy exponent.

To summarize, we believe we have attained a double goal.  Firstly, we 
have shown that models on $\phi^{3}$ graphs are good definitions of 
mean field spin glasses even where the dynamical behavior is 
concerned.  Secondly, we have determined with good numerical precision 
the exponents of the power decay of magnetization and energy.  This 
opens the way to a more systematic use of $\phi^{3}$ graphs for 
performing a numerical analysis of spin glasses.

\bigskip

\centerline{\bf ACKNOWLEDGEMENTS}

\medskip

We thank G. Parisi for interesting comments.

The bulk of the simulations were carried out
on the Front Range Consortium's
208-node Intel Paragon located at NOAA/FSL in Boulder.
CFB is supported by DOE under
contract DE-FG02-91ER40672 and by NSF Grand Challenge Applications
Group Grant ASC-9217394.
CFB and DAJ were partially supported by NATO grant CRG-951253.
DAJ, CN and EM are also supported in part by EC HCM
network grant CHRXCT930343.

\clearpage\newpage


\begin{table}
\begin{center}
\begin{tabular}{||r|c|r||}
\hline\hline
$\frac{T}{T_{c}}$ & time window & $\mu$ \\ \hline
$1/2$ & $20-2500$ & $0.37$ \\ \hline
$5/9$ & $20-2500$ & $0.39$ \\ \hline
$5/8$ & $20-1250$ & $0.43$ \\ \hline
$2/3$ & $10- 625$ & $0.46$ \\ \hline
$5/7$ & $10- 315$ & $0.49$ \\ \hline
$5/6$ & $10- 160$ & $0.54$ \\ \hline
$1$ & $ 5-  77$ & $0.67$ \\ \hline
\hline
\end{tabular}
\end{center}
\caption[2]{Magnetization time decay exponent $\mu$ for different $T$
values. $N=4000$. We also show the time window used for determining the
exponent.}
\protect\label{T-TAB1}
\end{table}

\vskip1cm

\begin{table}
\begin{center}
\begin{tabular}{||r|r||}
\hline\hline
$\frac{T}{T_{c}}$  & $\epsilon$ \\ \hline
$1/2$ & $0.45$ \\ \hline
$5/9$ &  $0.50$ \\ \hline
$5/8$ &  $0.53$ \\ \hline
$2/3$ &  $0.58$ \\ \hline
$5/7$ &  $0.64$ \\ \hline
$5/6$ &  $0.75$ \\ \hline
$1$   &  $0.92$ \\ \hline
\hline
\end{tabular}
\end{center}
\caption[2]{Energy time decay exponent $\epsilon$ for different $T$
values. $N=4000$. }
\protect\label{T-TAB2}
\end{table}

\clearpage\newpage


\clearpage\newpage

\clearpage\newpage

\clearpage\newpage

\clearpage\newpage

\clearpage\newpage

\clearpage\newpage

\end{document}